\documentclass[aps,twocolumn,floatfix, showpacs, superscriptaddress]{revtex4-1}
\usepackage[pdftex]{graphicx}% Include figure files
 \usepackage{amsmath}
\usepackage{subfigure}
\usepackage[utf8x]{inputenc}
\usepackage[T1]{fontenc}
\usepackage{amssymb}
\usepackage{amsfonts}
\usepackage{bm}
 \usepackage{amsmath} 
 \usepackage[breaklinks=true,colorlinks=true,linkcolor=blue,urlcolor=blue,citecolor=blue]{hyperref}
\usepackage{subfigure}
\usepackage{lipsum}
\usepackage{amsfonts} 
\usepackage{amssymb, mathrsfs}
\usepackage{braket}
\usepackage{graphicx} 
\usepackage[usenames]{color} 

\usepackage{bbm}
\def\beq{\begin{equation}}
\def\eeq{\end{equation}}
\def\bsp{\begin{split}}
\def\esp{\end{split}}
\def\bea{\begin{eqnarray}}
\def\eea{\end{eqnarray}}
\def\ba{\begin{array}}
\def\ea{\end{array}}

\def\lb{\left(}
\def\rb{\right)}

\def\l.{\left.}
\def\r.{\right.}

\begin{document}

%\date{\today}
\title{A toy model for quantum spin Hall effect}
\author{S. A. Owerre}
\email{solomon.akaraka.owerre@umontreal.ca}
\affiliation{D\'epartement de physique,
Universit\'e de Montr\'eal,
Montr\'eal 
Qu\'ebec H3C 3J7, Canada. }
\affiliation{Perimeter Institute for Theoretical Physics, 31 Caroline St. N., Waterloo Ontario N2L 2Y5, Canada.}

\author{J. Nsofini}
\email{jnsofini@uwaterloo.ca}
\affiliation{Institute for Quantum Computing, University of Waterloo, Waterloo, Ontario N2L 3G1, Canada.}
\affiliation{Department of Physics and Astronomy, University of Waterloo, Waterloo, Ontario N2L 3G1, Canada.}

\begin{abstract}

In this Communication, we investigate a toy model of  three-dimensional topological insulator surface, coupled homogeneously to a fictitious pseudo spin-$\frac{1}{2}$ particle. We show that this toy model captures the interesting features of topological insulator surface states, which include topological quantum phase transition and quantum spin hall effect.  We further incorporate  an out-of-plane magnetic field and obtain the Landau levels. 
\end{abstract}


\maketitle

{\it Introduction}.--
In recent years, topological insulators (TIs) have captivated considerable attention  of researchers \cite{yu, hk, hk1, hk2, zhang, km}. These fascinating materials involve compounds such as Bi$_2$Se$_3$ or Bi$_2$Te$_3$,  whose electronic bulk structure is an insulator with a finite gap separating the conduction band and the valence band, but their edges (for 2D TIs) or surfaces (for 3D TIs) have gapless states, which are protected by time-reversal symmetry $(TRS)$. These states are robust to perturbations that do not break this symmetry. In recent years, these gapless states  have been the focus of interest due to the simple Dirac-like Hamiltonian which describes them. However, breaking $TRS$ introduces many interesting phenomena; this can be achieved by  depositing a ferromagnet or a superconductor  on the surface of a TI.  Thus, the topologically protected surface state develops a gap, which leads to interesting electronic transport, which include half quantized conductivity, Majorana bound states, etc.\cite{yu, bur1, franz, sol3, dim, kane, zhang1, eze}. They also have potential technological applications in the field of spintronics\cite{zhang2, take}.

 In this Communication, we study a toy model that captures some of the interesting features of surface states in topological insulators.  Specifically,  we study  a fictitious pseudo spin-$\frac{1}{2}$ particle interacting homogeneously with the surface of a TI.  Assuming anisotropic ``in-plane'' and ``out-of-plane'' exchange interactions on the interface, say $\lambda_{\parallel}$ and $\lambda_{\perp}$,  the coupled system exhibits a topological  quantum phase transition as a function of $\xi=\lambda_{\perp}/\lambda_{\parallel}$.  
%\begin{figure}[ht]
%\centering
%\includegraphics[width=2in]{topo}
%\caption{Color online. Schematic diagram of the system considered in this paper. A nanoparticle (e.g SMM with giant spin $\boldsymbol{\mathcal{S}}$) in form of a quantum dot (red spot) is deposited on the surface of a $3D$ topological insulator.}
%\label{topo}
%\end{figure}
This phase transition is  associated with a quantum phase transition point at $\xi=1$, which  separates two regions; $\xi<1$, with two topologically protected Dirac points (semimetallic phase) and $\xi>1$, which is fully gapped (quantized spin Hall phase). 
We further show that when the $z$-component of the pseudospin is conserved, the system decouples into two Hamiltonians which are related by $TRS$.   The coupling term opens two gaps at the $\Gamma$ point on the interface, which are degenerate but with opposite spin orientations. Each gap opening gives rise to a half quantized Hall conductivity. Upon applying an electric field, an opposite spin current is induced on each interface leading to a quantized spin Hall conductivity.   Furthermore, we introduce a non-conserving term in the quantized spin Hall Hamiltonian by breaking the conservation of the $z$-component of the pseudospins. This leads to the obliteration of the quantized spin Hall conductivity. However, in this non-conserving regime we obtain a nontrivial topological spin Chern number  by diagonalizing the projection of the pseudospin onto the occupied valence bands \cite{pro}.

%The oscillatory motion of the mechanical resonator can be describe by the harmonic oscillator Hamiltonian
%\bea
%H_R=\frac{L_z}{2I_z}+\frac{I_z\omega_R^2\theta^2}{2}
%\eea
%where $L_z=-i\hbar\partial_\theta$ is the angular momentum operator.
 {\it Simplest Single Dirac Point}.-- Most of the interesting physics of  $2D$ metals  are captured by investigating the robustness of the band touching points (nodes). The bulk energy band near these points usually replicates a $2D$ Dirac Hamiltonian; thus these nodes are called Dirac points. The simplest form of Dirac point can be mapped out from this simple Hamiltonian: 
\bea
H=v_F(\hat{z}\times \boldsymbol{\sigma})\cdot\bold{k},
\label{8}
\eea
where $v_F$ is the Fermi velocity, $\sigma_i, i=x,y,z$ are the  Pauli matrices representing the real spins of the surface states. The $z$-component is taken to be perpendicular to the plane of the $2$D sample, and $\bold k=(k_x,k_y)$ is the $2$D surface Brillouin zone momentum. This Hamiltonian describes many known physical systems, such as the surface of a $3$D topological insulator \cite{zhang}, and the Dirac point Hamiltonian in graphene in which the Pauli matrices are pseudospins \cite{net0}. Evidently,  this Hamiltonian is  invariant when the spin and momentum go to minus of their original values; in other words, the Hamiltonian possesses $TRS$, where the time reversal operator is given $\Theta= i\sigma_y\mathcal{K}$;  $\mathcal{K}$ is a complex conjugation.   The energy spectrum is trivial and given by $\mathcal{E}_s=sv_F\sqrt{k_x^2+k_y^2}$; $s=\pm $. It is apparent that the two bands touch each other at $\bold k=0$. The robustness of this Dirac point is guaranteed by any perturbation that preserves the $TRS$ of Eq.\eqref{8}. By breaking $TRS$, the degeneracy of the band might be lifted and many interesting phenomena can emerged. There are several terms that break $TRS$; for instance,  $\Delta_x\sigma_x$ and $\Delta_y\sigma_y$ do break $TRS$; they correspond to Zeeman magnetic  field contributions along the $x$ and $y$ directions respectively. However, addition of these terms to Eq.\eqref{8} simply shifts the position of the Dirac points to  $\bold k=(0, -\Delta_x/v_F)$ and $\bold k=(\Delta_y/v_F, 0)$ respectively. Thus, the degeneracy is not lifted. The only $TRS$ breaking term that lifts the degeneracy of the bands is the direct coupling of the surface electron spins to the Zeeman magnetic field term perpendicular to the plane, {\it i.e.,} $\Delta_z\sigma_z$. This term can also be generated by depositing a ferromagnet on  the surface electrons \cite{franz, kane, sol3}. It is evident that this contribution opens a gap of size $2|\Delta_z|$ at $\bold k =0$. Consequently, when the Fermi energy lies between the gap, this  leads to a half-quantized Hall conductivity \cite{res, zhang}. In graphene, however, there are an even number of Dirac points. In this case, the charge Hall conductivity vanishes in the ordinary insulating state \cite{neil, adm, hk}.

%  In the presence of a constant magnetic field along the $z$-direction, there are two contributions to the effective Hamiltonian. The first one is the magnetic field experienced by  the surface electrons through the Peierls substitution $\bold{k}\to \bold{k} +\frac{e}{c}\bold{A}$, also known as the orbital effects on the electrons. $\bold{A}=-yB\hat{x}$ is the vector potential for the Landau gauge.  The second contribution is a direct coupling of the surface electron spins to the magnetic field, {\it i.e}, the Zeeman term $\Delta_z\sigma_z$, where $\Delta_z=g_e\mu_B$, and $g_e$  the effective electron gyromagnetic constant.

 {\it Toy Model}.--
 As mentioned in the preceding sections, the surface of a $3$D topological insulator possesses many interesting features when $TRS$ is explicitly broken.  
In this section, we will consider a $3$D topological insulator whose surface is  homogeneously coupled to a fictitious pseudospin-$\frac{1}{2}$ particle; the low-energy effective Hamiltonian can be written  as
\begin{align}
\mathcal{H}=v_F(\hat{z}\times \boldsymbol{\sigma})\cdot\bold{k}-\lambda_{\parallel}\tau_x\sigma_x -\lambda_{\perp}\tau_z\sigma_z,
\label{appen1}
\end{align}
where  $\sigma_i$ and $\tau_i$, ($i=x,y,z$) are the Pauli matrices acting on the topological insulator space and the fictitious pseudospin-$\frac{1}{2}$ particle space respectively. The coupling constants $\lambda_{\parallel}$ and $\lambda_{\perp}$ are the anisotropic  in-plane and out-of-plane homogeneous exchange interactions. In most cases of physical interest, such interaction is usually inhomogeneous, {\it i.e.}, the last two terms in Eq.\eqref{appen1} should contain a delta function. Thus,  Eq.\eqref{appen1} does not describe any known physical system. However, it is possible that it might be applicable to a pseudo quantum qubit,  but we are not interested in any specific system, instead we will consider it as a toy model that captures some of the interesting physics of TI surface states.  Although our model does not describe any known physical system, it possesses some features that are similar to other models that describe known systems. This is the main purpose of this Communication. We will consider Eq.\eqref{appen1} as the  basis of our investigation in this Communication.  It is apparent that Eq.\eqref{appen1} explicitly breaks $TRS$, {\it i.e }, $[\Theta, \mathcal{H}]\neq 0$, where $\Theta= \tau_x\otimes i\sigma_y\mathcal{K}$. Due to the fictitious pseudospin-$\frac{1}{2}$ particle, Eq.\eqref{appen1} is obviously a $4$-band model.  

 {\it Topological Quantum Phase Transition}.--  Most known systems such as graphene and thin film topological insulators possess topological quantum phase transition. This is manifested at the gap closing point which separates an ordinary insulator from a quantized Hall insulator \cite{phaset, burr3}. These phases are usually  distinguished by a topological invariant quantity called the  Chern number.  It is generally defined as \cite{vol}
\begin{align}
 \mathscr{C}= \frac{1}{24\pi^2}\epsilon_{\alpha\beta\gamma}\text{Tr}\int d^3k\mathcal{G}\partial_{k_{\alpha}}\mathcal{G}^{-1}\mathcal{G}\partial_{k_{\beta}}\mathcal{G}^{-1}\mathcal{G}\partial_{k_{\gamma}}\mathcal{G}^{-1},
\end{align}
where $\epsilon_{\alpha\beta\gamma}$ is the totally antisymmetric tensor, $\alpha= 0, x,y,z$, etc., labels the components of a four-vector,  and the  Matsubara Green's function can be written as
\bea
\mathcal{G}^{-1}(i\omega_n, \bold{k})= i\omega_n\mathbf{I}_{4\times 4}-\mathcal{H}.
\eea

\begin{figure}[ht]
\centering
\includegraphics[width=3.5in]{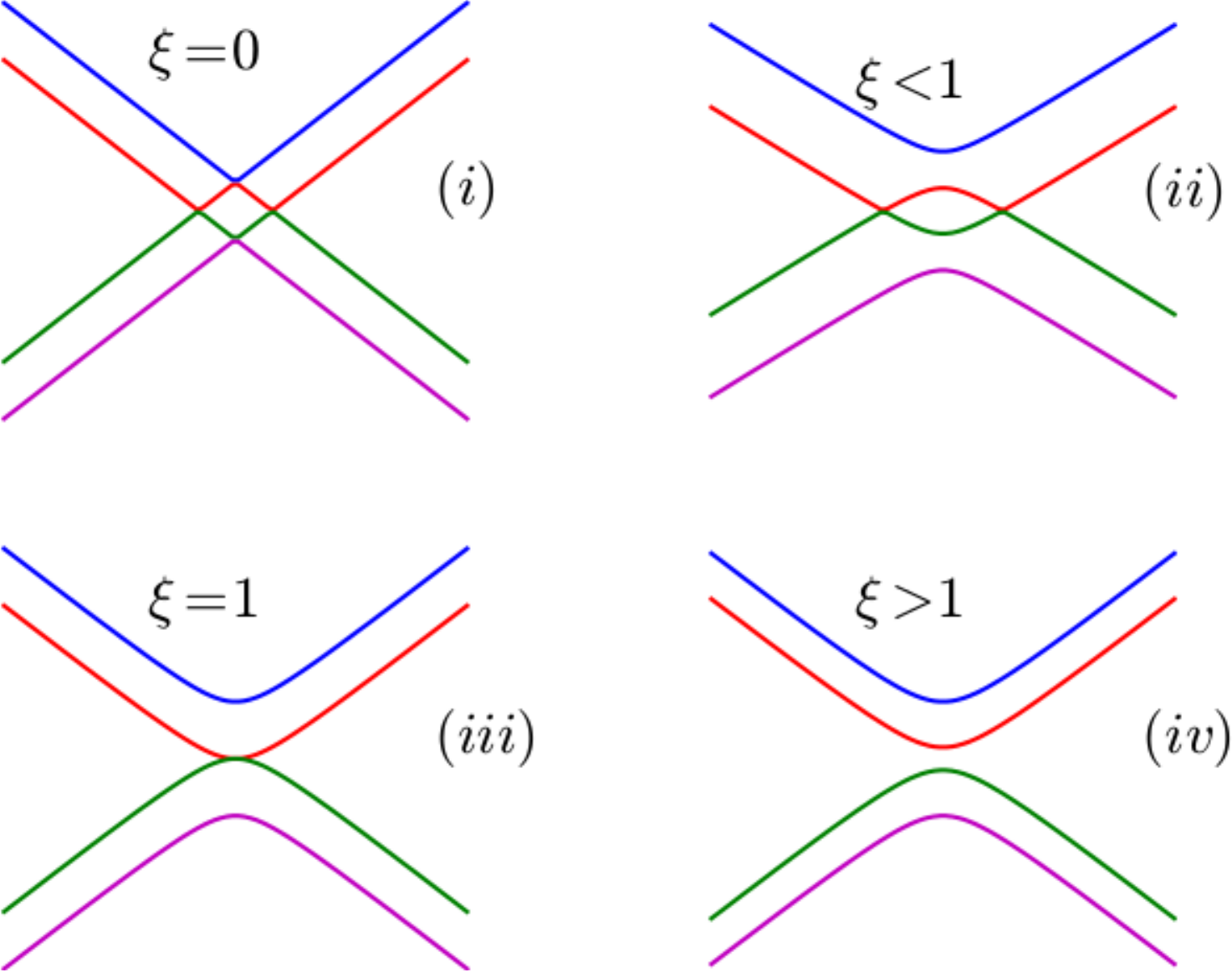}
\caption{Color online.  The energy bands along the $k_y$ direction for different ranges of the parameters; see text for explanation.}
\label{phase}
\end{figure}

The identity $\mathbf{I}_{4\times 4}$ is a $4\times 4$  matrix and $k_\alpha=(i\omega_n,\bold{k})$ is the momentum four-vector. In principle the Chern number can be computed for any model of interest.  One finds that it has a unique value in each phase, which is immutable by any smooth deformation of the system (provided the gap does not close). In general, the emergence of  quantum phase transition in  electron systems requires the violation of CPT symmetry \cite{vol}, {\it i.e.}, charge conjugation, parity, and time-reversal symmetries. It is evident that  our model in Eq.\eqref{appen1} explicitly breaks parity and time-reversal symmetries. Thus it can capture quantum phase transitions similar to those predicted in known systems. In order to see this, we diagonalize  Eq.\eqref{appen1}  
and find that the eigenvalues  are given by
\begin{align}
\mathcal{E}_{s\eta}=(-1)^{\eta}\sqrt{v_F^2k_x^2 +\lb \sqrt{v_F^2k_y^2 + \lambda_{\perp}^2}+ s\lambda_{\parallel} \rb^2},
\end{align}
where  $s=\pm$ and $\eta=0,1$. There are two energy sectors $\mathcal{E}_{+\eta}$ and $\mathcal{E}_{-\eta}$. In each sector there are two bands with $\eta=0$ being the conduction band and $\eta=1$ being the valence band. 
\begin{figure}[ht]
\centering
\includegraphics[width=2.5in]{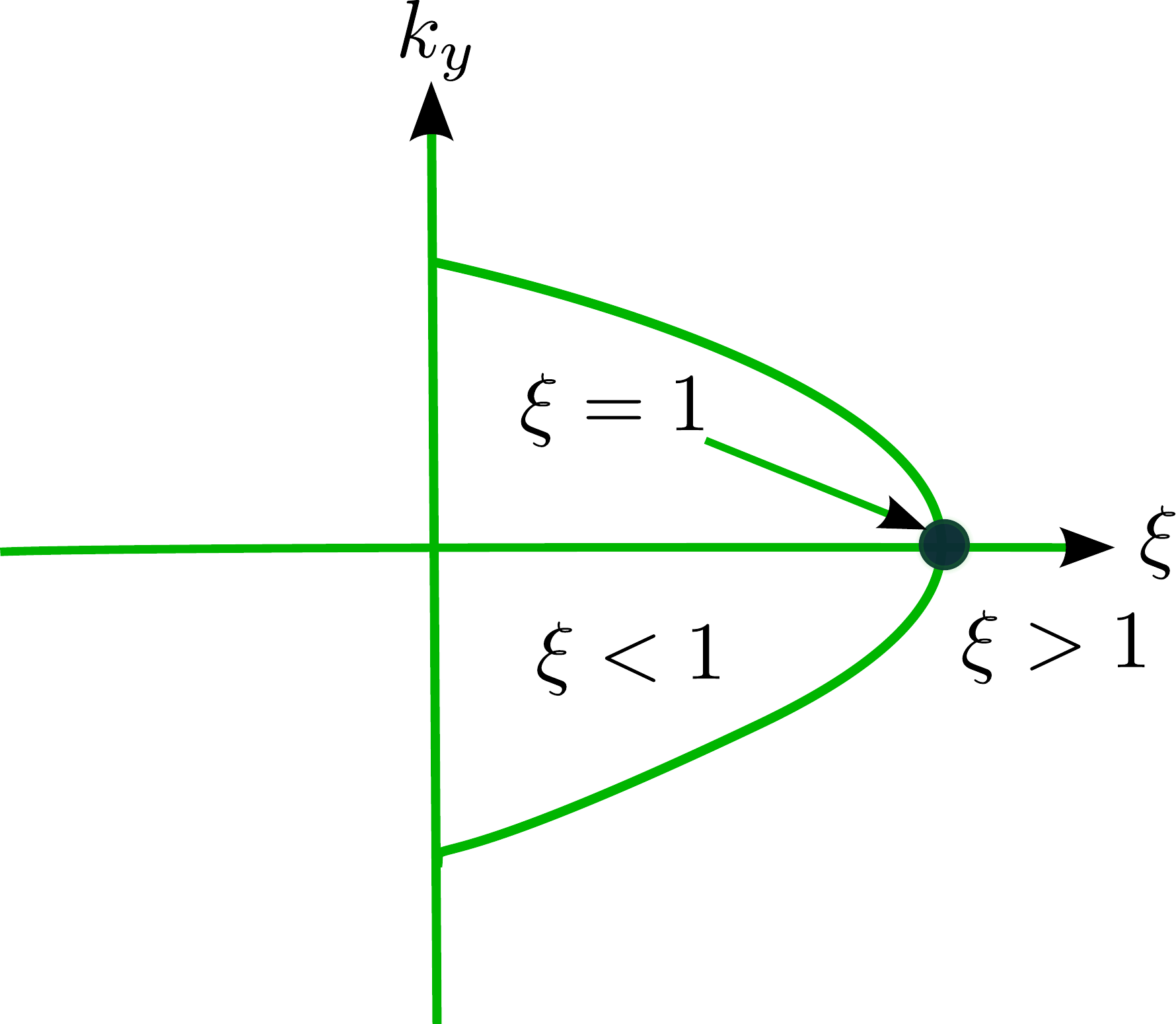}
\caption{Color online. The phase transition in $(\xi,k_y)$ space with $v_F=1=\lambda_{\parallel}$. The quantum phase transition point $\xi=1$  separates two regions; $\xi<1$, with two Dirac points and $\xi>1$, which is fully gapped; see Fig.\eqref{phase}.}
\label{Dirac}
\end{figure}
The $\mathcal{E}_{-\eta}$ bands have  two Dirac points ($\mathcal{E}_{+\eta}$ bands are always gapped) at $\pm\bold{K}$, where  
%It is evident that the  $\mathcal{E}_{+\eta}$ bands are always gapped, while the $\mathcal{E}_{-\eta}$ bands have two Dirac points given by
\bea
\bold{K}= \lb 0, \thinspace \frac{\lambda_{\parallel}}{v_F}\sqrt{1-\xi^2}\rb; \quad \xi=\lambda_{\perp}/\lambda_{\parallel}.
\label{wav}
\eea 
The topological quantum phase transition as a function of $\xi$ can be understood as follows. At the point $\xi = 0$, the system is a semimetal with two Dirac points separated by the wave vector $v_Fk_y= 2\lambda_{\parallel}$ along the $y$-direction. The semimetallic phase is controlled by the in-plane exchange interaction.  Provided $\xi <1$, these two Dirac points remain intact,  separated in momentum space by a wave vector $v_Fk_y= 2\lambda_{\parallel}\sqrt{1-\xi^2}$. They are topologically protected and cannot be eliminated by changing the parameters of the Hamiltonian. However, as $\xi$ is increased, the two Dirac points remain stable and subsequently annihilate each other at the phase transition point $\xi =1$, and emerge as a single Dirac point at the $\Gamma$ point $\bold{K}=(0,0)$. The gap further reopens as one moves to the region $\xi >1$ and the system becomes an insulator with a gap of $2\lambda_{\parallel}(\xi-1)$ at $\bold{K}=(0,0)$. As shown in Fig.\eqref{Dirac} and Fig.\eqref{phase}, the quantum phase transition point $\xi=1$  separates two regions; $\xi<1$, with two Dirac points and $\xi>1$, which is fully gapped. The homogeneous in-plane coupling $\lambda_{\parallel}$ plays a similar role as the parallel magnetic field in a thin film topological insulator \cite{burr3}.

{\it Quantum Spin Hall Effect}.-- Quantum spin Hall effect is known to be manifested in many known systems \cite{km,And,bat, hai1, hui1}. In this section we will show how the toy model in Eq.\eqref{appen1} captures this interesting physics. Let us consider the limit $\lambda_{\parallel}\to 0$. In this limit there are no Dirac points, the system is fully gapped. The Hamiltonian decouples into two copies which are related by $TRS$:
\begin{align}
\mathcal{H} =
 \begin{pmatrix}
  \mathcal{H}_{\uparrow} & 0  \\
 0 & \mathcal{H}_{\downarrow}
 \end{pmatrix},
 \label{eqn7}
 \end{align}
 where
 \bea
 \mathcal{H}_{\downarrow,\uparrow}= v_F(k_y\sigma_x-k_x\sigma_y) \pm\lambda\sigma_z,
 \label{ham}
 \eea
 and $\lambda=\lambda_\perp$.
 The eigenvalues are pair degenerate ($\tau_z$ is conserved),  given by 
 
 \begin{align}
\mathcal{E}_{s}&=s\sqrt{ v_F^2k^2 +\lambda^2}=s\epsilon_k,
\label{12} 
\end{align}
\begin{figure}[ht]
\centering
\includegraphics[width=2.5in]{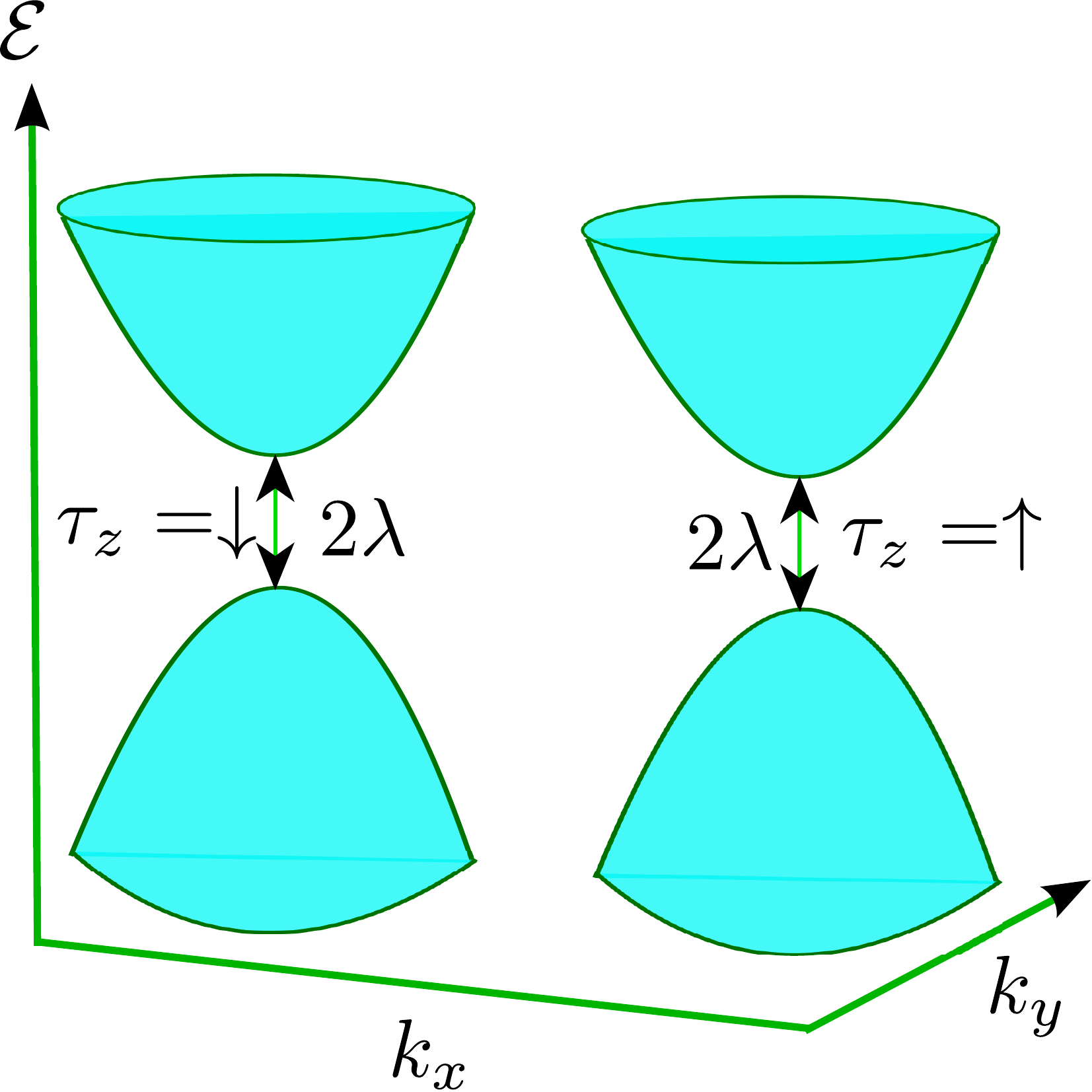}
\caption{Color online. Schematic diagram of the two Kramers energy bands on the interface with a gap of $2\lambda$ at the $\Gamma$ point.}
\label{cone}
\end{figure}
%\begin{figure}[ht]
%\centering
%\def\svgwidth{250pt}
%\input{cone1.pdf_tex}
%\caption{Color online. Schematic diagram of the two Kramers energy bands on the interface with a gap of $2\lambda$ at the $\Gamma$ point. } 
%\label{cone}
%\end{figure}
%
 where $k=\sqrt{k_x^2+k_y^2}$. The corresponding eigenspinors are found to be
 \begin{align}
 \psi_s^{\uparrow}&=  {\chi_s(\lambda) \choose 0} \quad\text{and} \quad \psi_s^{\downarrow}=  {0\choose \chi_s(-\lambda)},
  \end{align}
where $\chi_s(\lambda) = \frac{1}{\sqrt{2}}\lb\sqrt{1+s\frac{\lambda}{\epsilon_k}},-is e^{i\theta_k}\sqrt{1-s\frac{\lambda}{\epsilon_k}}\rb^T$ and $\theta_k =\text{tan}^{-1}\lb k_y/k_x \rb$. Fig.\eqref{cone} shows the Kramers pairs of the energy bands for spin-up and spin-down, with an  energy gap of $2\lambda$ at  the $\Gamma$ point $(k_x=0, k_y=0)$.  At zero temperature, and when the Fermi energy lies between the gap, the conduction band is empty  and the valence band is full; the Hall conductivity is given by \cite{thou}

\bea
\sigma_{xy}^{\tau_z}= \frac{e^2}{h}\mathscr{C}^{\tau_z},
\label{thou0}
\eea
where
\begin{equation}
 \mathscr{C}^{\tau_z}=\frac{1}{2\pi}\int d^2k\mathcal{F}_{\tau_z},
\label{thou1}
\end{equation}
is a topological invariant Chern number, $\mathcal{F}^{\tau_z}=\lb\boldsymbol{\nabla}\times \bold{A}^{\tau_z}\rb_z$ is the Berry curvature and $\bold{A}^{\tau_z}=i\braket{\psi_{-}^{\tau_z}|\boldsymbol{\nabla}|\psi_{-}^{\tau_z}}$ is the Berry connection. Expressing the differential operator in polar coordinates yields:
\bea
\mathcal{F}_{\tau_z}= \frac{1}{k}\lb\frac{\partial{(kA_{\theta_k})}}{\partial k}-\frac{\partial A_k}{\partial\theta_k}\rb,
\label{becu}
\eea
where $A_{\theta_k}=\frac{1}{k}i\braket{\psi^{\tau_z}_-|\partial_{\theta_k}\psi^{\tau_z}_-}$ and $A_{k}=i\braket{\psi^{\tau_z}_-|\partial_{k}\psi^{\tau_z}_-}$. Since $\theta_k$ appears as a $U(1)$ phase in the eigenspinors, $A_{\theta_k}$ and $A_k$ are independent of $\theta_k$, thus the second term in Eq.\eqref{becu} vanishes and the resulting expression becomes \cite{hui}:
\begin{equation}
\mathcal{F}^{\tau_z}= \frac{1}{k}\frac{\partial}{\partial k}\mathcal{M}^{\tau_z}(k); \quad \mathcal{M}^{\tau_z}(k)=i\braket{\psi^{\tau_z}_-|\partial_{\theta_k}\psi^{\tau_z}_-},
\label{thou2}
\end{equation}

Integrating Eq.\eqref{thou1} using Eq.\eqref{thou2} yields a half quantized Hall conductivity at each gap opening:
\bea
\sigma_{xy}^{\tau_z}= (\boldsymbol{\tau}\cdot\hat{z})\frac{-e^2}{2h}\text{sgn}(\lambda).
\label{hallc}
\eea
Evidently,  the Hall conductivity is coupled to the fictitious pseudospin orientations; see Fig.\eqref{cone}.  Thus, an application of an electric field will induce an opposite fictitious spin current. The charge-Hall conductivity vanishes due to $TRS$:
\bea
\sigma_{H}^c=\sigma_{xy}^{\uparrow}+\sigma_{xy}^{\downarrow}=0,
\eea
while the resulting  half quantized  spin Hall conductivity is nonzero:
\bea
\sigma_H^s=\frac{\hbar}{2e}\lb \sigma_{xy}^{\uparrow}-\sigma_{xy}^{\downarrow}\rb = -\frac{\text{sgn}(\lambda)}{2}\lb\frac{e}{2\pi}\rb.
\label{shf}
\eea
It should be noted that this spin Hall conductivity is not measurable. Physical known systems in which spin Hall effect has been realized  include  HgTe/CdTe semiconductor quantum wells \cite{bat}, and the ultrathin film topological insulator \cite{hai1}. Thus our coupling constant $\lambda$ plays a similar role as the coupling constant in these systems. 

{\it Effects of Non-Conserving Terms and Magnetic Field}.--
 Our analysis  for the  quantum spin Hall effect in the previous section relied on the fact that the $z$-component of the fictitious spin $1/2$ system is conserved. We will now investigate the effect of adding a non-conserving term to the Hamiltonian. There are many ways to do this, but we will choose the most plausible way. We assume that the fictitious particle is a two-level system with degenerate minima. Tunneling obviously breaks this degeneracy,  which introduces dynamics into the system and thus breaks the conservation of the $z$-component of the spin. This dynamics effectively corresponds to a term of the form $\Delta_x\tau_x$.  Thus, the model in Eq.\eqref{eqn7} then becomes:
 \begin{align}
\mathcal{H} =
 \begin{pmatrix}
  \mathcal{H}_{\uparrow} & \Delta_x  \\
 \Delta_x & \mathcal{H}_{\downarrow}
 \end{pmatrix}.
 \label{11a}
 \end{align}
 In this Hamiltonian, the off-diagonal term is akin to structure inversion asymmetry potential in a thin film topological insulator \cite{hui1}. % The system still preserves $TRS$ buts breaks parity.
%\begin{align}
%\mathcal{H}=v_F(\hat{z}\times \boldsymbol{\sigma})\cdot\bold{k}-\Delta_x\tau_x-\lambda\tau_z\sigma_z.
%\label{11a}
%\end{align}
Diagonalizing this Hamiltonian  we find that the eigenvalues are given by
\begin{align}
\mathcal{E}_{s\eta}&=(-1)^{\eta}\sqrt{\lb v_Fk +s\Delta_x\rb^2 +\lambda^2}.
\label{12a} 
\end{align}
 
% , as shown in Fig.\eqref{smm_split} . 
 % 
 The corresponding eigenspinors for  the conduction band ($\eta=0$) and the valence band ($\eta=1$) are given respectively by
\begin{align}
\chi_{s}^{\eta=0}=\lb -is e^{-i\theta_k} u_{s\eta +}, -su_{s\eta-},-ie^{-i\theta_k}u_{s\eta-}, u_{s\eta+}\rb^{T},
\end{align}
\begin{align}
\chi_{s}^{\eta=1}=\lb -is e^{-i\theta_k} u_{s\eta -}, -su_{s\eta+},ie^{-i\theta_k}u_{s\eta+}, u_{s\eta-}\rb^{T},
\label{vale}
\end{align}
where
\bea
u_{s\eta\pm}= \frac{1}{2}\sqrt{1\pm\frac{\lambda}{\mathcal{E}_{s\eta}}}.
\eea

\begin{figure}[ht]
\centering
\includegraphics[width=3.5in]{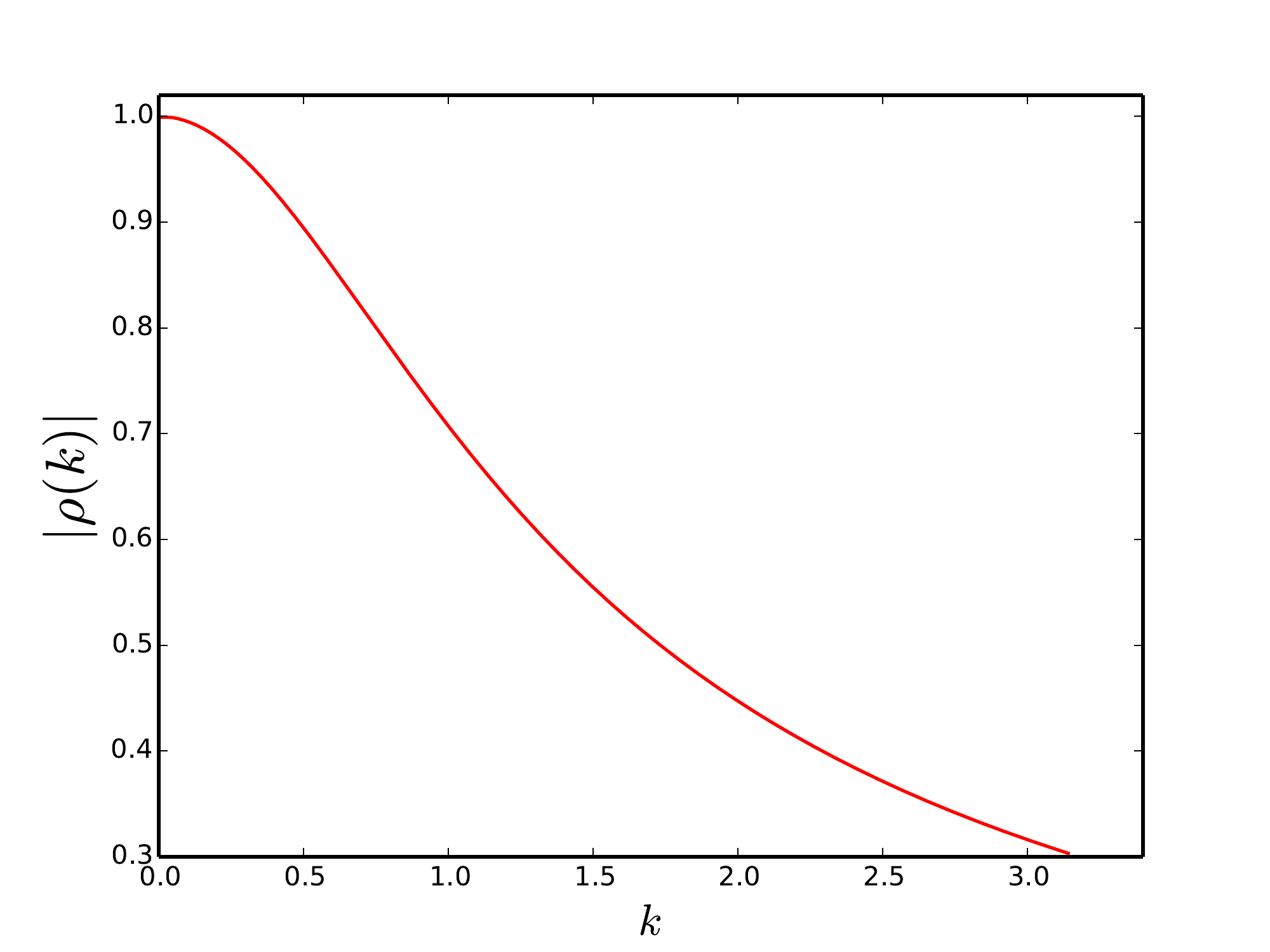}
\caption{Color online.  Plot of $\rho(k)$ vs. $k$ with $\Delta_x/\lambda =0.04$.}
\label{smm_split}
\end{figure}

%\section{Effects of magnetic field--Landau levels}
Since $\tau_z$ is no longer conserved, the Hall conductivity in Eq.\eqref{hallc} is obliterated. However, in the regime $\Delta_x/\lambda \ll 1$, one can still define a pseudo spin Chern number by diagonalizing the operator  $\mathcal{P}\tau_z\mathcal{P}$ \cite{pro, hui}, where $\mathcal{P}=\sum_{s=\pm}\chi_{s}\chi^{\dagger}_s$ is the projection operator onto the occupied valence bands in Eq.\eqref{vale}.  The matrix elements are expressed as $\braket{\chi_{s}|\tau_z\otimes\mathbf{I}_\sigma|\chi_{s^{\prime}}}$; $s,s^{\prime}=\pm$,  and the eigenvalues of this matrix are given by $\omega_{\pm}(k)=\pm|\rho(k)|$, where 
\bea
\rho(k)=2\lb u_{++}u_{-+}-u_{--}u_{+-}\rb,
\eea
and $\eta=1$ always. The corresponding eigenfunctions of $\tau_z$ are given by
\bea
\Psi_{\pm}=\frac{1}{\sqrt{2}}\lb\chi_{+}\pm \chi_{-}\rb.
\eea
The pseudo spin Chern number is now given by Eqs.\eqref{thou1} and \eqref{thou2} by replacing $\psi^{\tau_z}_-$ with $\Psi_{\pm}$ in $\mathcal{M}(k)$. Computing the corresponding inner products yields:
\bea
\mathcal{M}_{\pm}(k) = \pm\frac{\rho(k)}{2} +1.
\eea
Integrating Eq.\eqref{thou1} one obtains the Chern number:
\begin{equation}
 \mathscr{C}^{\pm}=\mathcal{M}^{\pm}(\infty)-\mathcal{M}^{\pm}(0)=\pm\frac{1}{2}\lb \rho(\infty)-\rho(0)\rb,
\label{thou3}
\end{equation}
which defines a nontrivial topological spin Hall phase provided that  $\Delta_x/\lambda \ll 1$. The function $\rho(k)$ is depicted in  Fig.\eqref{smm_split} for a specific value of $\Delta_x/\lambda$.  It is analogous to the {\it Pfaffian} \cite{km, hui1}, which is obtained by diagonalizing the $TRS$ operator $\Theta$ in the occupied valence bands basis. 

Interesting features also arise in the presence of a magnetic field. Now let us apply a constant magnetic field along the out-of-plane direction in the system described by Eq.\eqref{11a}. The resulting Hamiltonian becomes:
\begin{align}
\mathcal{H}&=v_F(\hat{z}\times \boldsymbol{\sigma})\cdot \bold{k}_{\bold A}+(\Delta_z-\lambda\tau_z)\sigma_z-\Delta_x\tau_x,
\label{16}
\end{align}
where $\bold k_{\bold A}=(-i\boldsymbol{\nabla}+\frac{e}{c}\bold{A})$ is the  magnetic field experienced by  the surface electrons through the Peierls substitution, also known as the orbital effects on the electrons. Here $\bold{A}=-yB\hat{x}$ is the vector potential for the Landau gauge. 
Diagonalization of Eq.\eqref{16} is achieved by introducing the creation and annihilation operators:
\begin{align}
b= \frac{l_B}{\sqrt{2}}(\Pi_x+i\Pi_y); \quad b^{\dagger}= \frac{l_B}{\sqrt{2}}(\Pi_x-i\Pi_y),
\end{align}
where $l_B^2=c/eB$ is the magnetic length and $\boldsymbol{\Pi}=-i\boldsymbol{\nabla} +\frac{e}{c}\bold{A}$.
Thus, Eq.\eqref{16} becomes:
\begin{align}
\mathcal{H}=i\omega_B\sqrt{2}(\sigma^+b-\sigma^-b^{\dagger})+(\Delta_z-\lambda\tau_z)\sigma_z-\Delta_x\tau_x,
\label{18}
\end{align}
with $\omega_B=v_F/l_B$ being the characteristic frequency analogous to cyclotron frequency. 
The corresponding landau levels for $n\neq0$ are obtained as:
\begin{align}
\mathcal{E}^2_{n\pm}&=2\omega_B^2n+\Delta_x^2 +\Delta_z^2+\lambda^2\nonumber\\&\pm 2\sqrt{2\omega_B^2n\Delta_x^2+\Delta_z^2\lb\lambda^2+\Delta_x^2\rb},
\label{19a} 
\end{align}
 There are two energy sectors for each Landau level which are controlled by several parameters. Both sectors behave quite differently as a function of the magnetic field. The Hall conductivity can as well be computed using these Landau levels \cite{bur2,bur3}. 
 
 {\it Conclusion}.--
 In conclusion, we have studied a toy model that captures the interesting physics of quantum spin Hall effect and topological quantum phase transition on the surface of a 3D topological insulator. Our model is comprised of a homogeneous exchange coupling between the surface electrons and the fictitious spin-$\frac{1}{2}$ particle; we showed that this  model exhibits both quantum phase transition and half quantized spin Hall effect. We presented a lucid exposition of the quantum phase transition and the topological invariant quantity that characterizes the Dirac points.  Introducing dynamics in the fictitious particle leads to a non-conserving term in the Hamiltonian, which breaks the conservation of $z$-component; thus obliterates the spin Hall effect. We obtained a nontrivial topological spin Chern number by diagonalizing the projection of the pseudo spin onto the occupied valence bands. Furthermore, we obtained the Landau levels in the presence of a magnetic field. 
 
 {\it{ Acknowledgments}.--}
The authors would like to thank Tami Pereg-Barnea for useful discussions, and  Manu Paranjape for his continual supports. This work commenced at  Universit\'e de Montr\'eal,
 and was completed  at
Perimeter Institute for Theoretical Physics.
\end{document}